# Market Basket Analysis Using Rule-Based Algorithms and Data Mining Techniques

Marina Kholod
Plekhanov Russian University of Economics
Kholod.MV@rea.ru

Nikita Mokrenko
Plekhanov Russian University of Economics
Mokrenko.NV@rea.ru

*Abstract -* Customer loyalty remains a critical yet elusive aspect of modern business strategy in today's data-driven environment. The rapid expansion of digital databases offers opportunities to analyze consumer behavior, but the sheer volume of information often obscures actionable insights. This study leverages market basket analysis, a form of data mining, to examine point-of-sale (POS) transactions from convenience stores, uncovering hidden patterns in purchasing behavior. The research identifies association rules that can inform marketing strategies and enhance operational efficiency. A structured methodology is applied to extract and interpret meaningful relationships within transactional data, emphasizing their implications for managerial decision-making. By demonstrating the potential of data mining to transform raw data into valuable business insights, this paper provides a framework for using analytical tools to improve customer engagement and competitive positioning.

## I. INTRODUCTION

In today's competitive, customer-focused, and service-driven economy, many modern companies face the challenge of unpredictable customer loyalty. The rapid expansion of databases in recent years, fueled by the widespread adoption of information and communication technologies, has made it easier to collect customer data by tracking daily activities and preferences. However, as the volume of data grows, the percentage that is clearly understood diminishes. Despite this, the data contains crucial insights for addressing the issue of fluctuating customer loyalty. By analyzing the behavior of past customers, it is possible to identify characteristics that distinguish those likely to remain loyal from those who might switch. One major challenge for companies is extracting meaningful information from transactional databases, which are relatively inexpensive to create, to identify these behavior patterns. This approach can provide a competitive edge and serves as a prime example of data-driven marketing. The focus of this paper is on uncovering implicit, previously unknown, and potentially valuable insights into customer purchasing behavior from raw data.

By utilizing data mining techniques, businesses can fully leverage their data to uncover previously hidden trends, behaviors, and anomalies. One effective type of data analysis in marketing is market basket analysis, which aims to enhance sales strategies and overall effectiveness by using customer data already available to the organization. The core objective of market basket analysis is to uncover association rules, a key component of database marketing or knowledge discovery within databases.

The objective of this study is to perform a market basket analysis using point-of-sale (POS) data from convenience stores, providing insights into consumer shopping behavior.

The structure of this paper is as follows: Section 2 offers a concise review of the background on applying data mining techniques to predict consumer purchasing behavior. Section 3 outlines the methodology, implements the analysis, and discusses the results. Section 4 interprets the association rules, highlights how to identify valuable patterns, and presents managerial implications. Finally, Section 5 provides a summary and conclusion of the study.

## II. RELATED WORKS AND BACKGROUND

In the modern business landscape, prioritizing "customer share" – the extent of business conducted with individual customers – has proven more beneficial than focusing solely on market share. Establishing personalized, one-to-one relationships with customers is therefore a crucial strategy for organizations of all sizes. Customer Relationship Management (CRM) facilitates this approach by implementing reliable systems, processes, and practices for engaging with customers. CRM is not merely a software solution but a customer-centric business strategy.

CRM streamlines and integrates various customer service processes within a company, enhancing the overall customer experience, which sets it apart from traditional database marketing. The CRM application architecture is typically divided into three components:

1. Operational: Includes sales force automation, customer service and support, and enterprise marketing automation.
2. Collaborative/Co-operational: Encompasses communication channels, call centers, and transaction management.
3. Analytical: Focuses on business intelligence for evaluating consumer behavior.

This paper emphasizes the analytical aspect of CRM, specifically the role of data mining as a tool for business intelligence. Data mining involves exploring and analyzing vast datasets to identify meaningful patterns and rules. It is closely related to concepts like Knowledge Discovery in Databases (KDD), knowledge extraction, and pattern analysis.

In the context of CRM, data mining enables organizations to gain a deeper understanding of their customers, thereby enhancing marketing strategies and customer support. While data mining techniques have been in existence for years, their commercial applications have gained traction over the past decade due to advancements in technology and organizational practices.

These developments include:

1. Increased automation in daily operations, such as transaction processing systems, point-of-sale scanners, and automated teller machines.
2. The ability to warehouse, clean, and summarize data for analysis.
3. Access to powerful servers capable of large-scale data mining, supported by sophisticated software tools.

This paper explores how these technological and organizational advancements have contributed to the rising importance of data mining in CRM.

Previous studies on applying market basket analysis (MBA), using the cash receipt as the unit of analysis for marketing purposes, are relatively scarce. However, existing literature on MBA can be broadly classified into three categories: model-based and empirical research, machine learning studies, and database-focused literature.

The model-based approach, commonly associated with MBA, is often employed to study association structures. For example, Bocker (1978) and Merckle (1981) utilized cash receipts to evaluate interdependence between product categories, relying on a frequency matrix of simultaneous purchases for all product pairs. Due to the computational challenges involved, they introduced a key assumption that product sales exhibit symmetric and transitive relations.

Julander (1989, 1992) emphasized the advantages of analyzing market baskets instead of individual brand sales, drawing the following conclusions:

Market baskets represent shoppers' natural interaction with retail environments and reflect actual purchases made during store visits. This allows averages and distributions of shopper baskets to serve as statistical indicators of shopping behavior.

Using the proportion of shoppers purchasing a specific product and the total number of receipts, it becomes possible to estimate the number of shoppers visiting and buying from the store.

Given the lack of efficient tools to process market baskets at the time, Julander (1992) manually entered 955 cash receipts from a Swedish supermarket into a database to conduct MBA.

Dickinson et al. (1992) introduced metrics such as the Base Compatibility Index (BCI) and the Inherent Compatibility Index (ICI) to evaluate merchandise compatibility. These metrics measure the frequency of two items appearing together relative to their expected frequency under an assumption of independence.

Bultez et al. (1996) employed MBA to reorganize retail assortments based on in-store shopping behaviors, utilizing Yule's Q-coefficient to analyze category associations. Their goal was to form retail business units by clustering highly interdependent categories.

Chintagunta and Haldar (1995) applied bivariate hazard and probit models to measure cross-category dependence in purchase timing and choice, though their work focused exclusively on category pairs. Similarly, Hruschka (1991) proposed a probabilistic model using logit equations to evaluate cross-category dependence across 72 categories.

Agrawal and Sikant (1994) introduced a probabilistic approach using data mining algorithms to uncover association rules, highlighting frequently purchased category subsets.

Other studies have explored the impact of promotions on non-promoted products, particularly their complementary effects. Walters (1991) and Mulhern and Leone (1991) provided evidence of asymmetric promotion effects across category pairs. Schmalen and Pechtl (1995) examined cross-category promotional effects, such as the influence of coffee promotions on other product categories. Chintagunta and Haldar (1995) extended their earlier models by incorporating variables for price and promotions, revealing stronger interdependence measures compared to models without these variables.

Manchanda et al. (1997) analyzed multi-category purchases across four product groups (laundry detergents, fabric softeners, cake mix, and frosting) using a multivariate probit model, finding significant complementary price effects between laundry detergents and fabric softeners.

Hruschka et al. (1999) used a multivariate binomial logit model to study cross-category dependence and promotion effects in retail assortments based on MBA data. However, this model had limitations, including restricting analyses to category-level dependencies and considering only first-order interactions between categories due to the complexity of item-level calculations.

Vindevogel, Poel, and Wets (2005) examined promotional effects on sales using association analysis. While MBA is widely viewed as a tool for developing effective price-promotion strategies based on the assumption of positive cross-price elasticity among associated products, negative cross-price elasticity is also observed. For instance, "horizontal variety seeking" behavior, where consumers purchase substitutes within the same category during a single trip, can lead to associations between competing products. In such cases, price promotions may negatively affect associated product sales. Conversely, Poel, Schamphelaere, and Wets (2004) found that complementary products performed well under specific promotional scenarios.

### III. METHODOLOGY

This work extends from the work depicted in [5]. The data mining methodology involves 11 steps:

*Step 1:* Translating a business problem into a data mining problem

The goals of our analysis are:

1. To identify all products in retail stores that are likely to respond effectively to promotional activities,
2. To gain insights into customer purchasing behavior within retail stores.

These objectives can be accomplished using the data mining technique known as Affinity Grouping (Association Rules).

*Step 2*: Choosing the relevant data

The availability and usefulness of data sources vary depending on the specific problem and industry. For our analysis of consumer behavior in the retail industry, we selected POS data from convenience stores. To analyze existing customers, ideally, we would include data from the time of their acquisition, along with information about their current status and behavioral patterns. However, it is not always feasible to obtain all desired data. The data we selected includes examples of all potential outcomes of interest. Since our data mining techniques are undirected, the POS data is considered adequate for our purposes.

*Step 3*: Getting to Know the Data

We began by profiling the data to better understand its structure.

All categories are food-related, although some products, such as newspapers, lack category data.

Data visualization tools were instrumental in the initial exploration of the database. Figure 1 shows categorical data, revealing that 84.52% of customers are women and 15.48% are men.

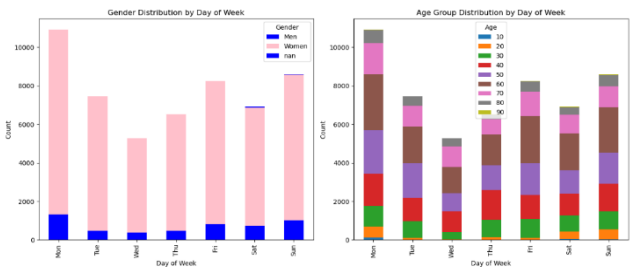

Figure 1. Categorical data distribution

At this stage, our goal is to explore the raw data for valuable insights, particularly focusing on the relationships between products. We aim to identify which products are strongly associated with one another and which are not. The MultiWebs visualization highlights the strength of these connections: thin lines indicate weak associations, while thick, solid lines represent strong connections, and medium-strength relationships are depicted with regular lines.

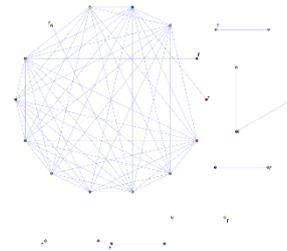

Figure 2. MultiWeb of Main Purchased Products

During the process of uncovering affinities within the basket contents, Figure 2 demonstrates the associations between product categories, with darker lines highlighting the strongest connections. By analyzing the interactions between individual items, clusters of related products can be identified, each consisting of multiple items that are independent from other sections of the assortment. Figure 2 shows a MultiWeb generated based on these interaction parameters.

*Step 4*: Creating a Model Set

The model set encompasses all the data used in the modeling process, so it is important to have data that spans different times of the year. Relying on data from a single period increases the risk of generating results that may not be universally applicable. In our case, we only have data for January, which follows the New Year holidays. This should not lead to unexpected results or false rules in the context of convenience stores.

*Step 5*: Addressing Data Issues

Data quality is a critical aspect of the data mining process, and there are common issues that can complicate the analysis, such as missing values. Missing data can arise from unknown, uncollected, or incorrectly entered information. Some modeling techniques handle missing values better than others. For example, A Priori and Generalized Rule Induction (GRI) are capable of managing missing values explicitly marked as "missing." Other techniques may struggle with missing data, resulting in longer training times and less accurate models. In our dataset, missing values occur in some categories, such as newspapers, which were excluded from the association analysis for the product categories.

Step 6: Transforming Data

Data transformation is an essential part of the mining process, though it is not always linear. While there is a general sequence, the steps can be revisited as needed. Data transformation involves defining data types and creating cross-tabulations. For A Priori, symbolic fields need to be transformed into flags or sets, while numeric fields should be converted into ranges. Most attributes in market basket data are zero for most instances, as customers typically purchase only a small fraction of the available items. This results in a sparse matrix, where rows represent customers and columns represent stock items, and most entries are zero.

Step 7: Building the Model

We use the following procedures for rule mining:

Standard Market Basket Analysis (A Priori and GRI):

Rule complexity depends on the number of items included. However, in convenience stores, customers tend to buy fewer items at a time, so looking for rules involving four or more items is impractical. In our analysis, we first apply A Priori and GRI, which are standard algorithms for large datasets. Prior to applying the algorithms, we define minimum support, confidence, and the number of items in the rule.

Market Basket Analysis Using Virtual Items:

To explore relationships between purchase behavior and other factors (such as gender, day of the week, and age), virtual item analysis is conducted. After completing the standard analysis, gender data is prepared and A Priori is applied again. The next step involves extracting data for each day of the week and age category before running the A Priori analysis.

Step 8: Testing Models

Model testing assesses whether the model works as expected. Evaluation can occur at both the model level and the level of individual predictions. In descriptive models, some rules are more informative than others. One way to measure the effectiveness of rules is through the minimum description length (MDL), which calculates the number of bits required to encode the rule and its exceptions. Fewer bits indicate better rules. Another method is the lift ratio, commonly used to assess affinity grouping models.

Validation is critical in data mining to ensure that the model generalizes to the population of interest. Typically, data mining methods do not rely on assumptions about data distribution (e.g., normality), so validation is essential. This is achieved by splitting the data: a portion is used to train the model, and predictions are tested on the remaining data. This helps verify that the model applies to data beyond that used to create it.

Step 9: Model Deployment

Deploying the model involves transferring it from the data mining environment to the scoring environment, where it will be used to evaluate large datasets. Implementing the model into the scoring platform is beyond the scope of this paper.

Step 10: Assessing Results

At this stage, results are evaluated based on business success rather than statistical criteria. Key performance indicators for the implemented model include profit and return on investment.

Step 11: Iteration

Data mining often raises more questions than it answers. When new relationships are discovered, new hypotheses can be tested, restarting the data mining process

## IV. RESULTS

The results of association analysis are presented as rules that describe the relationships between specific products or their categories. These rules are straightforward, and the strength of the product associations can be easily assessed. The outcome of market basket analysis typically yields three types of rules: actionable (providing valuable, usable information), trivial (already well-known to anyone familiar with the business), and inexplicable (offering no clear explanation or actionable insight). In the market basket analysis of convenience store data, several strong association rules were identified.

These results can be interpreted in various ways, depending on the decision maker's goals. This paper specifically focuses on brands with high purchase frequency. The most frequently purchased individual items include:

- Chokan Nikkan Sports Newspaper (1410 purchases)
- COCA-COLA Georgia Emerald Mountain Blend (1027 purchases)
- NISSHIN Cup Noodles (495 purchases).
- Table 1 lists the brands with the highest frequency, with a total of 1544 brands analyzed.

Table 1. Most Frequent Brands

| Brands | Frequency | % |
|---|---|---|
| Chokan Nikkan Sports Newspaper | 1410 | 15.26 |
| Coca-Cola Georgia Emerald Mountain Blend | 1027 | 11.11 |
| Nisshin Cup Noodles | 495 | 5.36 |
| Cool Delica Temaki Onigiri Shake | 79 | 0.85 |
| Cool Delica Temaki Onigiri Kombu | 69 | 0.75 |
| Chokan Sankei Sports Newspaper | 64 | 0.69 |
| Yamazaki DY Nikuman | 62 | 0.67 |
| Chokan Sports Newspaper | 56 | 0.61 |

Given the large number of generated rules, it can be challenging to determine which ones are actionable in a convenience store setting and which are not. Ultimately, this decision will depend on the goals and interpretation of the decision-maker. As previously mentioned, all three measures—support, confidence, and lift—are crucial. Therefore, additional rules that maximize each of these measures will be presented. These selected rules can help identify the most significant patterns and relationships in the data, making it easier to prioritize actions. By focusing on the rules with the highest support, confidence, and lift, we can uncover the most impactful insights for the business. A notable enhancement to the standard use of association rules is the concept of virtual items. Virtual items allow the analysis to incorporate additional information beyond individual products, leveraging data contained in the database. In our analysis, we consider aggregated transaction data, such as customer characteristics and the day and time of purchase, as virtual items. Including virtual items in the association analysis can uncover purchase patterns, such as "if 'Chokan Sports' and 'Saturday', then 'JT Frontier Menthol'." The table below illustrates some of these rules.

Table 2. Market Basket Analysis of Virtual Items (days of the week, gender, age)

| ID | Consequent | Antecedent | Instances | Support % | Confidence % | Lift |
|---|---|---|---|---|---|---|
| 1 | JT Frontier Menthol Box | Chokan Sports Newspaper SATURDAY | 6 | 0 206 | 50 | 4865 |
| 2 | Newspaper | Yukan Sports WEDNESDAY | 6 | 0 206 | 50 | 4865 |
| 3 | Newspaper | Coca-Cola WEDNESDAY | 6 | 0 206 | 50 | 4865 |
| 4 | Newspaper | Chokan WEDNESDAY | 6 | 0 206 | 50 | 4865 |
| 5 | JT Frontier Menthol Box | Chokan Chokan 30"s | 6 | 0 206 | 50 | 4865 |
| 6 | JT Frontier Menthol Box | Chokan MEN | 6 | 0 206 | 50 | 4865 |
| 7 | Natori Yaki | Asahi Super Dry Nama Chokan Nikkan Sports 20"s | 6 | 0 206 | 50 | 4865 |
| 8 | JT Frontier Menthol Box | Chokan Newspaper Chokan Chokan Nikkan | 6 | 0 206 | 50 | 4865 |
| 9 | JT Frontier Menthol Box | Chokan Sports Newspaper Chokan MEN 30"s | 6 | 0 206 | 50 | 4865 |
| 10 | Natori Yaki Aji | Asahi Super Dry Nama Chokan MEN 20"s | 6 | 0 206 | 50 | 4865 |
| 11 | JT Frontier Menthol Box | Chokan Chokan Chokan MEN 30"s | 6 | 0 206 | 50 | 4865 |

## V. CONCLUSION

This paper explores the use of association rule analysis, including virtual items, to uncover purchasing patterns in convenience store data. Virtual items enable the integration of additional information, such as customer and store characteristics, to identify differences in purchase behavior across consumer types. While market basket mining effectively reveals consumer patterns, challenges include managing large datasets and assessing the strength of numerous generated rules.

The paper emphasizes the importance of measures like support, confidence, and lift in evaluating rule quality, while also highlighting the limitations of the analysis, such as reliance on POS data, the need for large transaction volumes, and the risk of overfitting results from data limited to a single time period. Despite these challenges, the analysis provides valuable insights into consumer profiles and product affinities in convenience stores.

## FUNDING

This research was funded by the Ministry of Science and Higher Education of the Russian Federation, grant number FSSW-2023-0004.